\documentstyle[iopconf1,epsfig]{article}

\gdef\Feynmanlength{\setlength{\unitlength}{0.01pt}}  

\global\newcount\LINETYPE
\global\newcount\LINEDIRECTION
\global\newcount\LINECONFIGURATION
\newcommand{\LTYPE}{\LINETYPE}
\newcommand{\LDIR}{\LINEDIRECTION}

\global\LINETYPE=1  \global\LINEDIRECTION=0  \global\LINECONFIGURATION=0
\global\newcount\fermion    \fermion=1
\global\newcount\scalar     \scalar=2
\global\newcount\photon     \photon=3
\global\newcount\gluon      \gluon=4
\global\newcount\SPECIAL    \SPECIAL=5
\gdef\N{0}  \gdef\NE{1}  \gdef\E{2}   \gdef\SE{3}
       
\global\newcount\REG            \global\REG=0
\global\newcount\FLIPPED        \global\FLIPPED=1
\global\newcount\CURLY          \global\CURLY=2
\global\newcount\FLIPPEDCURLY   \global\FLIPPEDCURLY=3
\global\newcount\FLAT           \global\FLAT=4
\global\newcount\FLIPPEDFLAT    \global\FLIPPEDFLAT=5
\global\newcount\CENTRAL        \global\CENTRAL=6
\global\newcount\FLIPPEDCENTRAL \global\FLIPPEDCENTRAL=7
             
\global\newcount\SQUASHEDGLUON  \global\SQUASHEDGLUON=8

\newcount\adjx \adjx=0
\newcount\adjy \adjy=0
\global\newdimen\BIGPHOTONS     \BIGPHOTONS=0pt  
\global\newdimen\THICKPHOTONS     \THICKPHOTONS=0pt  
\global\newdimen\THICKPHOTONSWITCH    \THICKPHOTONSWITCH=0pt
\gdef\THICKPHOTONTEST{
\THICKPHOTONSWITCH=0pt
\ifdim\THICKPHOTONS=0pt \relax
  \else \ifnum\LTYPE=3
           \ifnum\LDIR=2 \THICKPHOTONSWITCH=1pt \fi 
           \ifnum\LDIR=6 \THICKPHOTONSWITCH=1pt \fi 
        \fi
\fi
}  
\gdef\THICKLINES{\thicklines  \THICKPHOTONS=1pt}

\global\newcount\phantomswitch   \global\phantomswitch=0
\global\newcount\stemlength   \global\stemlength=275   
\global\newcount\absstemlength        
\global\newcount\stemlengthx          
\global\newcount\stemlengthy          
\newdimen\FRONTSTEM  \FRONTSTEM=0pt   
\newdimen\BACKSTEM   \BACKSTEM=0pt    
\newdimen\EITHERSTEM \EITHERSTEM=0pt  
\global\newcount\arrowlength                
\global\newdimen\ATTIP   \global\ATTIP=0pt  
\global\newdimen\ATBASE  \global\ATBASE=1pt 
\global\newcount\unitboxnumber  
\global\newcount\unitboxnumberpo  
\global\newcount\particlelengthx  
\gdef\plengthx{\particlelengthx}
\global\newcount\particlelengthy  
\gdef\plengthy{\particlelengthy}
\global\newcount\boxlengthx  
\global\newcount\boxlengthy  
\global\newcount\particleadjustx  
\global\newcount\particleadjusty  
\global\newcount\particlelength   
\global\newcount\particlefrontx
\gdef\pfrontx{\particlefrontx}
\global\newcount\PFRONTx
\global\newcount\particlefronty
\gdef\pfronty{\particlefronty}
\global\newcount\PFRONTy
\global\newcount\particlebackx
\gdef\pbackx{\particlebackx}
\global\newcount\particlebacky
\gdef\pbacky{\particlebacky}
\global\newcount\particlemidx
\gdef\pmidx{\particlemidx}
\global\newcount\particlemidy
\gdef\pmidy{\particlemidy}
\global\newcount\seglength  \global\newcount\gaplength
\global\gaplength=850  
\global\seglength=1416  
\global\newcount\Xone    \global\newcount\Yone    
\global\newcount\Xtwo    \global\newcount\Ytwo    
\global\newcount\Xthree  \global\newcount\Ythree  
\global\newcount\Xfour   \global\newcount\Yfour   
\global\newcount\Xfive   \global\newcount\Yfive   
\global\newcount\Xsix    \global\newcount\Ysix    
\global\newcount\Xseven  \global\newcount\Yseven  
\global\newcount\Xeight  \global\newcount\Yeight  
\newsavebox{\lastline}  
\global\newcount\numlineparts   
\global\newcount\upperlineadjx  \upperlineadjx=0  
\global\newcount\upperlineadjy  \upperlineadjy=0  
\global\newcount\lowerlineadjx  \lowerlineadjx=0  
\global\newcount\lowerlineadjy  \lowerlineadjy=0  
\global\newcount\thirdlineadjx  \thirdlineadjx=0  
\global\newcount\thirdlineadjy  \thirdlineadjy=0  
\global\newcount\fourthlineadjx \fourthlineadjx=0  
\global\newcount\fourthlineadjy \fourthlineadjy=0  
\global\newcount\unitboxwidth   \unitboxwidth=1000
\global\newcount\unitboxheight  \unitboxheight=0  
\global\newcount\numupperunits  \numupperunits=8  
\global\newcount\numlowerunits  \numlowerunits=8  
\global\newcount\numthirdunits  \numthirdunits=8  
\global\newcount\numfourthunits \numfourthunits=8  
\global\newcount\fermioncount   \global\fermioncount=0
\global\newcount\scalarcount    \global\scalarcount=0
\global\newcount\photoncount    \global\photoncount=0
\global\newcount\gluoncount     \global\gluoncount=0
\global\newcount\SPECIALcount   \global\SPECIALcount=0
\global\newcount\vertexcount    \global\vertexcount=-1
\global\newcount\XDIR
\global\newcount\YDIR
\gdef\SETDIR{  
\ifcase\LDIR
     \global\XDIR=0  \global\YDIR=1   
\or  \global\XDIR=1  \global\YDIR=1   
\or  \global\XDIR=1  \global\YDIR=0   
\or  \global\XDIR=1  \global\YDIR=-1  
\or  \global\XDIR=0  \global\YDIR=-1  
\or  \global\XDIR=-1 \global\YDIR=-1  
\or  \global\XDIR=-1 \global\YDIR=0   
\or  \global\XDIR=-1 \global\YDIR=1   
\else\DIRECTERROR
\fi}  
\gdef\moduloeight#1{
\ifnum#1>7 \global\advance #1 by -8
\relax
\moduloeight#1
\relax
\else \relax
\fi}
\gdef\multroothalf#1{\global\multiply #1 by 7071 \global\divide #1 by 10000}
\gdef\negate#1{\global\multiply #1 by -1}

\gdef\slanttest(#1,#2){
\ifodd\LDIR
\multiply #1 by 7071  \divide #1 by 10000
\multiply #2 by 7071  \divide #2 by 10000
\fi
}
\gdef\gslanttest(#1,#2){
\ifodd\LDIR
\multroothalf#1
\multroothalf#2
\fi
}
\gdef\setplength{ 
\global\particlelengthx=\unitboxwidth
\global\particlelengthy=\unitboxheight
\global\multiply \particlelengthx by \unitboxnumber
\global\multiply \particlelengthy by \unitboxnumber
\global\advance \particlelengthx by \particleadjustx
\global\advance \particlelengthy by \particleadjusty
}
\gdef\boxlengthdefault{  
\global\boxlengthx=\plengthx
\global\boxlengthy=\plengthy
\ifnum\plengthx<0 \global\multiply\boxlengthx by -1 \fi
\ifnum\plengthy<0 \global\multiply\boxlengthy by -1 \fi
}
\gdef\rearcoords{  
\global\particlebacky=\particlefronty
\global\particlebackx=\particlefrontx
\global\advance \particlebackx by \particlelengthx
\global\advance \particlebacky by \particlelengthy
}
\gdef\midcoords{  
\global\particlemidy=\particlefronty
\global\particlemidx=\particlefrontx
\global\stemlengthx=\particlelengthx  
\global\stemlengthy=\particlelengthy
\global\divide\stemlengthx by 2
\global\divide\stemlengthy by 2
\global\advance \particlemidx by \stemlengthx
\global\advance \particlemidy by \stemlengthy
}
\gdef\setparticle{\setplength\rearcoords\midcoords\boxlengthdefault}  
\gdef\setcoords(#1,#2,#3)(#4,#5,#6)[#7,#8]{
\global\upperlineadjx=#1
\global\lowerlineadjx=#2
\global\thirdlineadjx=#3
\global\upperlineadjy=#4
\global\lowerlineadjy=#5
\global\thirdlineadjy=#6
\global\unitboxwidth=#7
\global\unitboxheight=#8
}
\gdef\drawoldpic#1(#2,#3){  
\global\particlefrontx=#2
\global\particlefronty=#3
\rearcoords
\midcoords
\put(#2,#3){\usebox{#1}}
}
\gdef\drawsavedline`#1' as #2[#3#4](#5,#6)[#7]{
\global\LINETYPE=#2
\global\LINEDIRECTION=#3
\global\LINECONFIGURATION=#4
\global\particlefrontx=#5
\global\particlefronty=#6
\global\unitboxnumber=#7
\selectcase
\rearcoords
\midcoords
\ifnum\phantomswitch=0 \drawas{#1}\fi
}

\gdef\drawas#1{
\global\savebox{#1}(\boxlengthx,\boxlengthy){
\setlength{\unitlength}{0.01pt}
\begin{picture}(\boxlengthx,\boxlengthy)
\multiput(\upperlineadjx,\upperlineadjy)(\unitboxwidth,\unitboxheight)
{\numupperunits}{\upperunitbox}
\ifnum\numlineparts > 1  
\multiput(\lowerlineadjx,\lowerlineadjy)(\unitboxwidth,\unitboxheight)
{\numlowerunits}{\lowerunitbox}
\fi
\ifnum\numlineparts > 2  
\multiput(\thirdlineadjx,\thirdlineadjy)(\unitboxwidth,\unitboxheight)
{\numthirdunits}{\thirdunitbox}
\fi
\ifnum\numlineparts > 3  
\multiput(\fourthlineadjx,\fourthlineadjy)(\unitboxwidth,\unitboxheight)
{\numfourthunits}{\lowerunitbox}
\fi
\end{picture} }
\global\PFRONTx=\pfrontx  \global\PFRONTy=\pfronty   
\SETFRONTSTEM
\THICKPHOTONTEST
\ifdim\THICKPHOTONSWITCH=1pt\global\advance\PFRONTy by 20  \fi
\put(\PFRONTx,\PFRONTy) {\usebox{#1}}   
\ifdim\THICKPHOTONSWITCH=1pt
\global\advance\PFRONTy by -40
\put(\PFRONTx,\PFRONTy) {\usebox{#1}}   
\global\advance \PFRONTy by 20  
\fi  
\SETBACKSTEM
\seglength=1416   \gaplength=850   
}
\gdef\drawandsaveline`#1' as #2[#3#4](#5,#6)[#7]{
\global\newsavebox{#1}
\drawsavedline`#1' as #2[#3#4](#5,#6)[#7]
}

\gdef\drawline#1[#2#3](#4,#5)[#6]{   
\drawsavedline`\lastline' as #1[#2#3](#4,#5)[#6]}

\gdef\TYPEERROR{\message{*** ERROR IN PARTICLE TYPE SELECTION ***}
\message{+++ Try with line type \fermion,\scalar,\photon,\gluon
(see manual) +++}\SETERR}
\gdef\DIRECTERROR{\SETERR\message{*** ERROR IN PARTICLE DIRECTION SELECTION
***}
\message{+++ Try again with direction N, NE, E, SE  etc. or see manual +++}}
\gdef\UNIMPERROR{\message{*** ERROR IN PARTICLE OPTIONS SELECTION ***}
\message{
+++ The requested options combination has not yet been implemented +++}\SETERR}
\gdef\SETERR{\gdef\upperunitbox{{\tiny Error}}  
\gdef\lowerunitbox{\relax}
\gdef\thirdunitbox{\relax}
}
\gdef\neglengthcheck{\ifnum\unitboxnumber < 1
\message{   *** ERROR:  PARTICLE OF NEGATIVE OR ZERO LENGTH REQUESTED. ***   }
\message{   ***         TAKING ABSOLUTE VALUE. ***   }\negate\unitboxnumber
\fi}
\gdef\selectcase{
\neglengthcheck   
\SETDIR
\ifcase\LINETYPE
\TYPEERROR  
\or \selectfermion  
\or \selectscalar   
\or \selectphoton   
\or \selectgluon    
\or \selectspecial  
\else \TYPEERROR \fi  }
\gdef\selectfermion{
\ifnum\fermioncount=0 
\global\newcount\fermionlength  
\global\newcount\fermionlengthx
\global\newcount\fermionlengthy
\global\newcount\fermionfrontx  
\global\newcount\fermionfronty  
\global\newcount\fermionbackx
\global\newcount\fermionbacky
\gdef\ALLfermion{  
\global\fermionfrontx=\particlefrontx \global\fermionfronty=\particlefronty
\ifnum\unitboxnumber > 50000
\message{   *** WARNING *** Fermion of length
\the\unitboxnumber\space requested ***   }
\ifnum\unitboxnumber > 80000
\message{   *** Reducing fermion length to 30000 (max 80000) ***   }
\global\unitboxnumber=30000 \fi \fi  
\global\fermionlength=\unitboxnumber 
\global\particleadjustx=0   \global\particleadjusty=0 
\global\numlineparts = 1    \global\numupperunits=1
\global\upperlineadjx=-200  \global\upperlineadjy=0
\global\fermionlengthx=\fermionlength    \global\fermionlengthy=\fermionlength
\gslanttest(\fermionlengthx,\fermionlengthy)  
\global\multiply\fermionlengthx by \XDIR  
\global\multiply\fermionlengthy by \YDIR  
\global\unitboxheight=\fermionlengthy   \global\unitboxwidth=\fermionlengthx
\global\advance \fermionlengthx by \particleadjustx
\global\advance \fermionlengthy by \particleadjusty
\global\particlelengthx=\fermionlengthx
\global\particlelengthy=\fermionlengthy
\boxlengthdefault    \rearcoords    \midcoords
\global\fermionbackx=\particlebackx     \global\fermionbacky=\particlebacky
\ifcase\LINECONFIGURATION  
\ifnum\XDIR=0
\gdef\upperunitbox{\line(\XDIR,\YDIR){\boxlengthy}} 
\else
\gdef\upperunitbox{\line(\XDIR,\YDIR){\boxlengthx}}
\fi
\else \UNIMPERROR
\fi
}

 \fi
\global\advance\fermioncount by 1  
\ALLfermion
}
\gdef\selectscalar{
\ifnum\scalarcount=0 
\newcount\scalarlength
\newcount\scalarlengthx
\newcount\scalarlengthy
\newcount\scalarfrontx  
\newcount\scalarfronty  
\newcount\scalarbackx
\newcount\scalarbacky
\gdef\ALLscalar{
\global\scalarfrontx=\particlefrontx   
\global\scalarfronty=\particlefronty   
\numlineparts = 1      \numupperunits=\unitboxnumber
\ifcase\LINECONFIGURATION
\global\upperlineadjx=-200     \global\upperlineadjy=0
\slanttest(\seglength,\gaplength)   
\gdef\upperunitbox{\line(\XDIR,\YDIR){\seglength}}
\else \UNIMPERROR 
\fi
\global\unitboxwidth=\seglength  \global\advance\unitboxwidth by \gaplength
\global\multiply \unitboxwidth by \XDIR
\global\unitboxheight=\seglength  \global\advance\unitboxheight by \gaplength
\global\multiply \unitboxheight by \YDIR
\global\particleadjustx=\gaplength \global\multiply\particleadjustx by \XDIR
\global\particleadjusty=\gaplength \global\multiply\particleadjusty by \YDIR
\negate\particleadjustx   \negate\particleadjusty   
\setparticle  
\global\scalarlengthx=\particlelengthx  
\global\scalarlengthy=\particlelengthy  
\ifnum\boxlengthx > 50000
\message{   *** WARNING *** Scalar of length in excess of 50000cp
requested!}\fi
\ifnum\boxlengthy > 50000
\message{   *** WARNING *** Scalar of length in excess of 50000cp
requested!}\fi
\global\scalarbackx=\pbackx      \global\scalarbacky=\pbacky   
}

 \fi
\global\advance\scalarcount by 1  
\ALLscalar
}
\gdef\selectphoton{   
\ifnum\photoncount=0 \input PHOTONSETUP  \fi
\selectphoton
}
\gdef\selectgluon{   
\ifnum\gluoncount=0 \input GLUONSETUP  \fi
\selectgluon
}
\gdef\selectspecial{\UNIMPERROR}
\gdef\checkvertex{ 
\ifnum\vertexcount=-1   \input VERTEX  \fi}
\gdef\drawvertex#1[#2#3](#4,#5)[#6]{\checkvertex\drawvertex#1[#2#3](#4,#5)[#6]}
\gdef\vertexcap#1{\checkvertex\vertexcap#1}
\gdef\vertexcaps{\checkvertex\vertexcaps}
\gdef\vertexlink#1{\checkvertex\vertexlink#1}
\gdef\vertexlinks{\checkvertex\vertexlinks}
\gdef\stemvertex#1{\checkvertex\stemvertex#1}
\gdef\stemvertices{\checkvertex\stemvertices}
\gdef\flipvertex{\checkvertex\flipvertex}
\global\arrowlength=349  
\gdef\drawarrow[#1#2](#3,#4){
\global\LDIR=#1
\SETDIR
\global\boxlengthx=#3  
\global\boxlengthy=#4  
\ifdim#2=1pt  
\adjx=\arrowlength      \adjy=\arrowlength
\multiply\adjx by \XDIR \multiply\adjy by \YDIR  
\slanttest(\adjx,\adjy)
\global\advance\boxlengthx by \adjx    \global\advance\boxlengthy by \adjy
\fi
\ifnum\phantomswitch=0\put(\boxlengthx,\boxlengthy){\vector(\XDIR,\YDIR){0}}\fi
}  
\gdef\SETFRONTSTEM{
\EITHERSTEM=\FRONTSTEM   \advance\EITHERSTEM by \BACKSTEM
\ifdim\EITHERSTEM>0pt
\global\stemlengthx=\stemlength   \global\stemlengthy=\stemlength
\global\absstemlength=\stemlength
\SETDIR
\gslanttest(\stemlengthx,\stemlengthy)
\gslanttest(\absstemlength,\REG)  
\ifnum\XDIR=0 \stemlengthx=0 \fi
\ifnum\YDIR=0 \stemlengthy=0 \fi
\global\multiply\stemlengthx by \XDIR
\global\multiply\stemlengthy by \YDIR
\ifdim\FRONTSTEM=1pt
\ifnum\phantomswitch=0
          \put(\pfrontx,\pfronty){\line(\XDIR,\YDIR){\absstemlength}}\fi
\global\advance\plengthx by \stemlengthx
\global\advance\plengthy by \stemlengthy
\global\advance\PFRONTx by \stemlengthx
\global\advance\PFRONTy by \stemlengthy
\global\advance\pmidx by \stemlengthx
\global\advance\pmidy by \stemlengthy
\global\advance\pbackx by \stemlengthx
\global\advance\pbacky by \stemlengthy
\ifnum\LTYPE=3
\global\photonfrontx=\PFRONTx  \global\photonfronty=\PFRONTy
\global\photonbackx=\pbackx    \global\photonbacky=\pbacky
\fi  
\ifnum\LTYPE=4
\global\gluonfrontx=\PFRONTx  \global\gluonfronty=\PFRONTy
\global\gluonbackx=\pbackx    \global\gluonbacky=\pbacky
\fi  
\fi  
\fi  
}    
\gdef\SETBACKSTEM{
\ifdim\BACKSTEM=1pt
\ifnum\phantomswitch=0
       \put(\pbackx,\pbacky){\line(\XDIR,\YDIR){\absstemlength}}\fi
\global\advance\plengthx by \stemlengthx
\global\advance\plengthy by \stemlengthy
\global\advance\pbackx by \stemlengthx
\global\advance\pbacky by \stemlengthy
\fi  
\global\stemlength=275  \FRONTSTEM=0pt  \BACKSTEM=0pt 
}    
\gdef\drawloop#1[#2#3](#4,#5){  
\input LOOPS  
\drawloop#1[#2#3](#4,#5)}
\Feynmanlength  

\begin{document}
\title{Agut Masses}
\author{C D Froggatt\dag\ and
H B Nielsen\ddag}
\affil{\dag\  Department of Physics and Astronomy,
University of Glasgow, Glasgow G12 8QQ, UK}
\affil{\ddag\ Niels Bohr Institute, Blegdamsvej 17, Copenhagen $\phi$,
Denmark}

\beginabstract
{}From considerations of the number of matrix elements of
different orders of magnitude in the quark and charged
lepton mass matrices, we suggest that the underlying
gauge group responsible for the spectrum
should have several---actually of the
order of 7 to 10---cross product factors.
This is taken to support our AGUT gauge group $SMG^3 \times U(1)_f$
which, under certain simple conditions, is the maximal group
transforming the known 45 Weyl quark-lepton fields into each other.
We describe the AGUT fit to the charged fermion mass spectrum,
and briefly discuss baryogenesis and the neutrino mass problem.

\endabstract

\section{ Introduction}
What is the origin of the well-known pattern of large ratios
between the quark and lepton masses and of the small quark
mixing angles? This is the problem of the hierarchy of
Yukawa couplings in the Standard Model (SM).
We suggest \cite{fn} that the natural resolution to this
problem is the existence of some approximately conserved chiral
charges beyond the SM. These charges, which we assume to be
gauged, provide selection rules forbidding the transitions
between the various left-handed and right-handed fermion
states (except for the top quark).

For example, we suppose that there exists some charge (or
charges) $Q$ for which the quantum number difference between
left- and right-handed Weyl states is larger for the
electron than for muon:
\begin{equation}
\left| Q_{eL} - Q_{eR} \right| > \left|
Q_{\mu L} - Q_{\mu R} \right|
\end{equation}
It then follows that the SM Yukawa coupling for the electron
$g_e$ is suppressed more than that for the muon $g_{\mu}$,
when $Q$ is taken to be approximately conserved. This is
what is required if we want to explain the electron-muon
mass ratio.

In section 2 we give arguments motivating our identification
of the above chiral gauge charges with those of the
anti-grand unification theory (AGUT) based on the non-simple
gauge group $SMG^3 \times U(1)_f$, where
$SMG \equiv SU(3) \times SU(2) \times U(1)$. We also give
a crude statistical argument for the number of cross-product
factors in the gauge group beyond the SM, suggested by the
observed fermion spectrum. In section 3 we discuss the
structure of the AGUT gauge group and how it can be
rather simply characterized, as the maximal gauge group
satisfying a few simple principles. The Higgs fields
responsible for breaking the AGUT gauge group
$SMG^3 \times U(1)_f$ to the diagonal $SMG$ subgroup,
identified as the SM gauge group, are considered in
section 4. The structure of the resulting fermion mass
matrices are presented in section 5, together with
details of a fit to the charged fermion spectrum. In
sections 6 and 7, we briefly discuss the problems of
baryogenesis and neutrino oscillations respectively.
Finally we present our conclusions in section 8.

\section{Motivation for Anti-GUT}
As pointed out in the introduction, the quark-lepton mass
matrices---written in a basis of what we can call
proto-flavours---have matrix elements typically suppressed
relative to the electroweak scale ($<\phi_{WS}> = 246$ GeV)
by rather large factors. We shall take the point of view
that, in the fundamental theory beyond the SM,
the Yukawa couplings allowed by gauge invariance
are all of order unity and, similarly,
all the mass terms allowed by gauge invariance are of
order the fundamental mass scale of the theory---say
the Planck scale. Then, apart from the matrix element
responsible for the top quark mass, the quark-lepton
mass matrix elements are only non-zero due to the
presence of other Higgs fields having vacuum expectation
values (VEVs) smaller (typically by one order of magnitude)
than the fundamental scale. These Higgs fields will,
of course, be responsible for breaking the fundamental
gauge group $G$ down to the SM group. In order to generate
a particular effective SM Yukawa coupling matrix element,
it is necessary to break the symmetry group $G$ by a
combination of Higgs fields with the appropriate
quantum number combination  $\Delta \vec{Q}$. When this
``$\Delta \vec{Q}$'' is different for two matrix elements,
they will typically deviate by a large factor.
If we want
to explain the observed spectrum of quarks and leptons in this
way, it is clear that we need charges which---possibly in a
complicated way---separate the generations and, at least
for $t-b$ and $c-s$, also quarks in the same generation.
Just using the usual simple $SU(5)$ GUT charges does not
help, because both ($\mu_R$ and $e_R$) and
($\mu_L$ and $e_L$) have the same $SU(5)$ quantum numbers.
So we prefer to keep each SM irreducible representation
in a separate irreducible representation of $G$ and
introduce extra gauge quantum numbers distinguishing
the generations, by adding extra cross-product factors to
the SM gauge group.

In order to be guided into a specific model and to convince
ourselves that something like the model we propose should
represent nature, we should like to estimate how complicated
the gauge group $G$
should be.
In principle we could estimate over
how many different orders of magnitude
the various matrix elements in the
mass matrices should distribute themeselves,
because these matrix elements are to a large extent
accessible to phenomenological---almost
experimental---measurement. We could then adjust the
degree of complication of the group - say
the number of cross product factors
in the gauge group $G$ that should be used.

We consider the three charged particle
mass matrices - since we have
only rather uncertain information on the neutrinos anyway.
With 3 generations they clearly contain
together $3 \times 9 = 27$ different matrix elements.
The number of matrix elements that are essentially
measurable in practice is estimated as ca 12,
corresponding to the meaurement of $3 \times 3$ masses
and three mixing angles.  How many different order
of magnitude classes
of matrix elements then needed
will, of course, depend somewhat
on the attitude as to when a couple of estimated
matix elements  deviate by more than of ``order of
magnitude unity'', but we take roughly
the number
of classes to be around 7.
We get this
number 7 by saying that, of the ``observable'' 12 matrix elements
in our model to be presented below, we have used 8 different orders
of magnitude, but that there is
one clear case in which
there are used two different orders of magnitude in
principle, although the experimental numbers give that
the two matrix elements have the same order of magnitude:
The matrix element giving the 2nd to 3rd generation mixing
has, see eq.(\ref{Y_D}), the order $T^3$, while the one dominating
the b-quark mass is of order  $WT^2$, whereas they
are experimentally numerically almost equal.
If now we could guess that the non-observed
matrix elements would distribute themselves
into classes with the same order of magnitude
in much the same way---i.e. with much the same
number of matrix elements in the same
class---then we would come to
around $\frac{7}{12}  \times 27 \approx 16$ classes for
the elements in all three
matrices. This number, in succession, must now depend on how
many combination possibilities there are for the breaking of
the symmetries, in various ways, by means of the Higgs fields.
Obviously there should be more classes
the more pieces the gauge group consists of. The
connection between the number of cross product factors n,
say in the gauge group $G =$ $ A_1 \times A_2 \times \cdots
\times A_n $, and the number of order of magnitude
classes of mass matrix elements may be seen
crudely estimated in fig.~1.

\begin{figure}
\leavevmode
\centerline{\epsfig{file=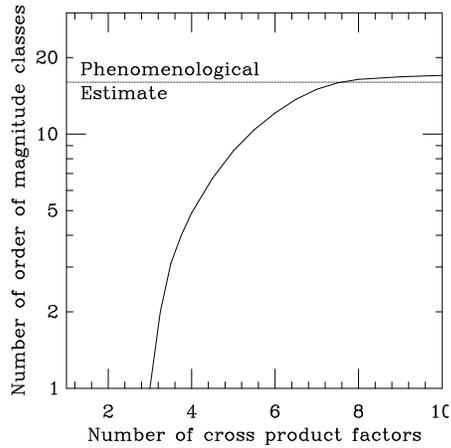,width=7.0cm}
}
\caption{The connection between the number of cross product
factors in the extended group $G$ and the number of order
of magnitude classes of quark-lepton mass matrix elements}
\label{fig:classes}
\end{figure}

This figure was constructed using three examples
and a hand drawn curve through the three points,
based on a vague expectation of how
it should be curved. The examples we chose were
the SM itself, the maximal AGUT group model
and the reduced AGUT model (in which one
of the Higgs fields $S$ has a VEV of order
unity in fundamental units)
which we actually use to fit the experimental
data. In the pure SM there are no suppression
factors, and all 27 mass matrix elements
are expected to belong to the same class
and be of order the electroweak
scale. Thus the SM corresponds to the point with the
number of order of magnitude classes equal to one and
number of cross product factors $n = 3$ on fig.~1.
The other end of the curve corresponds to the maximal
AGUT group $SMG^3 \times U(1)_f$, having $n = 10$ cross
product factors. In this case the matrix elements are
much more separated into different order of magnitude
classes. As we shall see below the corresponding
diagonal elements in each of the 3 Yukawa coupling
matrices $Y_U$, $Y_D$ and $Y_E$ have
the same order of magnitude. Furthermore we have
$(Y_U)_{12} \simeq (Y_D)_{12}$ and
$(Y_D)_{32} \simeq (Y_E)_{32}$,
giving a total of 8 constraints and
hence $27 - 6 - 2 = 19$ in
principal different order of magnitude classes.
If we take into account a couple of extra
numerical coincidences in order of magnitude,
the number of classes is reduced to 17.
The third point on the curve corresponds to the AGUT model, in
which the maximal AGUT group is replaced by the group to
which it is broken by the non-suppressing Higgs field $S$,
namely $SMG^2\times U(1)$. The latter group has 7 factors in the
cross product, and the reduced Yukawa matrices satisfy 2 extra
constraints: $(Y_U)_{12} = (Y_U)_{21}$ and
$(Y_D)_{12} = (Y_E)_{21}$. Thus the number of different order of
magnitude classes in this case is $17 - 2 = 15$.

The idea now is to estimate the number of cross product
factors in the ``true'' gauge group underlying
the fermion mass spectrum,
by looking up in figure 1 what
number of factors would give the number of
order of magnitude classes
found phenemenologically. We get about 7 to 8 factors
being suggested, not surprisingly since we used
our model with seven cross product factors
and the fact that it fits rather well to
determine one point on the curve. In spite of this
way of constructing the curve and what we consider the
phenomenological number, we do not consider it
an empty excercise to use our curve to estimate
with what accuracy we must expect that a
true model would have to have around the 7
cross product factors. Our point, of course, is to argue
that only by a rather strange accident
should it be possible for
models, with a very different number of cross
product factors, to fit the experimental data.
In particular if the model had significantly less than
7 factors, there should be too many mass matrix
elements not separated by the orders of magnitude
found experimentally. It is anyway not possible
to have more cross product factors than in the
maximal AGUT model
($n = 10$) and still have an anomaly free group.

At the end of the day, of course, the real
motivation for considering the rather specific
AGUT gauge group close to the Planck scale, with
order unity fundamental Yukawa couplings,
is its success in fitting the
masses of the quarks and leptons and the
quark mixing angles. This fit
is discussed in section 5.

\section{The ``maximal'' AGUT gauge group}

The AGUT model is based on extending the SM gauge group
$SMG = S(U(2) \times U(3))$ in a similar way to grand
unified $SU(5)$, but rather to the non-simple
$SMG^3 \times U(1)_f$ group. Here we follow
Michel and O'Raifeartaigh \cite{Michel}, in
using not only the Lie algebra
but even the Lie {\em group} for a
Yang-Mills theory. The global properties of the
$SMG$ group imply that the SM particle
representations of the $SU(3) \times SU(2) \times U(1)$
Lie algebra should obey the charge quantisation rule
\begin{equation}
\frac{y}{2} + \frac{1}{2}{\rm ``duality"} +
	\frac{1}{3}{\rm ``triality"} \equiv 0 \pmod 1
\label{SMGiChQu}
\end{equation}
which expresses
the somewhat complicated way in which the electric charge
is quantised in the SM. Here $y$ is the conventional
weak hypercharge, and ``duality'' is 1 for the
fundamental representation of SU(2)
(the doublet) and 0 for the singlet.
Similarly ``triality'' is 1 for the
SU(3) triplet, $-1$ for the anti-triplet and
0 for the singlet.

The existence of the $SMG^3 \times U(1)_f$ group means
that, near the Planck scale, there are three sets of
SM-like gauge particles. Each set only couples to its
own proto-generation [e.g. the proto- $u$, $d$, $e$ and
$\nu_e$ particles], but not to the other two proto-generations
[e.g. the proto- $c$, $s$, $\mu$, $\nu_{\mu}$, $t$, $b$,
$\tau$ and $\nu_{\tau}$ particles]. There is also an extra
abelian $U(1)_f$ gauge boson, giving altogether
$3 \times 8 = 24$ gluons, $3 \times 3 = 9$ $W$'s and
$3 \times 1 + 1 =4$ abelian gauge bosons. The couplings
of the $SMG_i = S(U(2) \times U(3))_i \approx SU(3)_i
\times SU(2)_i \times U(1)_i$ group to the $i$'th
proto-generation are identical to those of the SM
group. Consequently we have a charge quantisation
rule, analogous to eq.~(\ref{SMGiChQu}), for each
of the three proto-generation weak hypercharge
quantum numbers $y_i$.

To first approximation---namely in the approximation
that the quark mixing angles $V_{us}$, $V_{cb}$,
$V_{ub}$ are small---we may ignore the prefix ``$proto$-''.
However we really introduce in our model some
``proto-fields'' characterized by their couplings
to the 37 gauge bosons of the $SMG^3 \times U(1)_f$
group. The physically observed $u$-quark, $d$-quark
etc. are then superpositions of the proto-quarks
(or proto-leptons), with the same named proto-particle
dominating. Actually there is one deviation from this
first approximation rule that proto-particles correspond
to the same named physical particles. In the AGUT fit to
the quark-lepton mass spectrum, discussed below,
we find that to first approximation the right-handed components
of the top and the charm quarks must be permuted:
\begin{equation}
c_{R \ \mbox{proto}} \approx t_{R \ \mbox{physical}} \qquad
t_{R \ \mbox{proto}} \approx c_{R \ \mbox{physical}}
\end{equation}
But for all the other components we have:
\begin{equation}
t_{L \ \mbox{proto}} \approx t_{L \ \mbox{physical}} \qquad
b_{R \ \mbox{proto}} \approx b_{R \ \mbox{physical}}
\end{equation}
and so on.

The AGUT group breaks
down an order of magnitude or so below the Planck
scale to the SM group, as the diagonal subgroup of
its $SMG^3$ subgroup.
The gauge coupling constants do not,
of course, unify, but their values have been successfully
calculated using the so-called multiple point
principle \cite{glasgowbrioni}.

At first sight, this $SMG^3 \times U(1)_f$ group with
its 37 generators seems to be just one among many
possible SM gauge group extensions. However, we shall
now argue it is not such an arbitrary choice, as it
can be uniquely specified by postulating 4 reasonable
requirements on the gauge group $G$ beyond the SM.
As a zeroth postulate, of course, we require
that the gauge group extension must contain the Standard Model
group as a subgroup $G \supseteq SMG$.
In addition it should obey the
following 4 postulates:

\begin{enumerate}
\item $G$ should transform the presently known (left-handed,
say) Weyl particles into each other, so that
$G \subseteq U(45)$. Here $U(45)$ is the group of all
unitary transformations of the 45 species of Weyl fields (3
generations with 15 in each) in the SM.
\item No anomalies, neither gauge nor mixed.
We assume that only straightforward anomaly
cancellation takes place and, as in the SM itself,
do not allow for a Green-Schwarz type anomaly
cancellation \cite{green-schwarz}.
\item The fifteen irreducible representations of Weyl fields
for the SM group remain irreducible under $G$. This is
the most arbitrary of our assumptions about $G$. It
is motivated by the observation that combining SM
irreducible representations into larger unified
representations introduces symmetry relations between
Yukawa coupling constants, whereas the particle spectrum
doess not exhibit any exact degeneracies (except
possibly for the case $m_b = m_{\tau}$). In fact
AGUT only gets the naive $SU(5)$ mass predictions as
order of magnitude relations:
$m_b \approx m_{\tau}$, $m_s \approx m_{\mu}$,
$m_d \approx m_e$.
\item $G$ is the maximal group satisfying the other 3
postulates. We already argued, in the previous section, that the
large number of order of magnitude classes of fermion
mass matrix elements indicates the need for a large
number of cross product factors in $G$.
\end{enumerate}

With these four postulates a somewhat cumbersome
calculation shows that,
modulo permutations of the various irreducible representations
in the Standard Model
fermion system, we are led to our gauge group
$SMG^3\times U(1)_f$.
Furthermore it shows that the SM group is embedded
as the diagonal subgroup of $SMG^3$, as in our AGUT model.
The most difficult part of
the calculation is to decide which
identifications of the abelian groups
have to be made in order to avoid anomalies, i.e. how
big a subgroup of $U(1)^{15}$  can avoid having anomalies
and be allowed in $G$.
In searching for the generators of an allowed subgroup, one
may expand them in terms of the generators for these
15 $U(1)$ 's and they have to obey some first order
(linear) relations for the coefficients,
in order to avoid anomalies
involving also the non-abelian or gravitational fields.
Also there are third order relations that have to be
satisfied, in order that there be no anomalies involving only
the subspace of abelian generators. It turns out that, with
the three proto-generations of fermions, there are
too many constraints to be solved with an abelian
subgroup of dimension higher than 4.
It is found that three of the allowed abelian generators in $G$
can be taken to be the 3 weak hypercharges, each defined to act
on only one proto-generation. After that choice
the scheme becomes so tight that, apart from
various rewritings and permutations
of the particle names, there is a unique fourth $U(1)$ allowed
and that is what we call $U(1)_f$. Several of the anomalies
involving this $U(1)_f$ are cancelled by assigning
equal and opposite values of the $U(1)_f$ charge to
the analogous particles belonging to second and
third proto-generations, while the
first proto-generation particles have just
zero charge \cite{davidson}.
In fact the $U(1)_f$ group does not couple to
the left-handed particles and the $U(1)_f$ quantum
numbers can be chosen as follows for the proto-states:
\begin{equation}
Q_f(\tau_R) = Q_f(b_R) = Q_f(c_R) = 1
\end{equation}
\begin{equation}
Q_f(\mu_R) = Q_f(s_R) = Q_f(t_R) = -1
\end{equation}

Thus the quantum numbers of the quarks and leptons
are uniquely determined in the AGUT model. However
we do have the freedom of choosing the gauge quantum
numbers of the Higgs fields responsible for the breaking
the $SMG^3 \times U(1)_f$ group down to the SM gauge
group. These quantum numbers are chosen with a view to
fitting the fermion mass and mixing angle data \cite{smg3m},
as discussed in the next section.

\section{Symmetry breaking by Higgs fields}

\label{choosinghiggs}

There are obviously many different ways to break down the
large group $G$ to the much smaller SMG. However, we can
first greatly simplify the situation by
assuming that, like the quark and lepton fields, the Higgs
fields belong to singlet or fundamental representations of
all non-abelian groups. The non-abelian representations are
then determined from the $U(1)_i$ weak hypercharge quantum
numbers, by imposing the charge quantisation rule
eq.~(\ref{SMGiChQu}) for each of the $SMG_i$ groups.
So now the four abelian charges, which we express in
the form of a charge vector
\begin{displaymath}
\vec{Q} = \left( \frac{y_1}{2}, \frac{y_2}{2},
\frac{y_3}{2}, Q_f \right)
\end{displaymath}
can be used to specify the complete representation of $G$.
The constraint, that we must eventually recover the SM
group as the diagonal subgroup of the $SMG_i$ groups,
is equivalent to the constraint that all the Higgs fields
(except for the Weinberg-Salam Higgs field which of course
finally breaks the SMG) should have charges $y_i$ satisfying:
\begin{equation}
\label{diagU1}
y=y_1+y_2+y_3=0
\end{equation}
in order that their SM weak hypercharge $y$ be zero.

We wish to choose the charges of the Weinberg-Salam (WS) Higgs
field, so that it matches the difference in charges between
the left-handed and right-handed physical top
quarks. This will ensure that the top quark
mass in the SM is not suppressed relative
to the WS Higgs field VEV. However,
as we remarked in the previous section, it is
necessary to associate the physical right-handed
top quark field not with the corresponding third
proto-generation field $t_R$, but rather with the right-handed
field $c_R$ of the second proto-generation. Otherwise
we cannot suppress the bottom quark and tau lepton masses.
This is because, for the proto-fields, the charge differences
between $t_L$ and $t_R$ are the same as between $b_L$
and $b_R$ and also between $\tau_L$ and $\tau_R$. So
now it is simple to calculate the quantum numbers of
the WS Higgs field $\phi_{WS}$:
\begin{equation}
\vec{Q}_{\phi_{WS}} = \vec{Q}_{c_R} - \vec{Q}_{t_L}
	= \left( 0,\frac{2}{3},0,1 \right) -
	\left( 0,0,\frac{1}{6},0 \right)
	= \left( 0,\frac{2}{3},-\frac{1}{6},1 \right)
\end{equation}
This means that the WS Higgs field
will in fact be coloured under both $SU(3)_2$ and
$SU(3)_3$. After breaking the symmetry down to the SM
group, we will be left with the usual WS Higgs field
of the SM and another scalar which will be an octet of
$SU(3)$ and a doublet of $SU(2)$.
This should not present any phenomenological problems,
provided this scalar doesn't cause symmetry breaking
and doesn't have a mass less than the few TeV scale.
In particular an octet of $SU(3)$ cannot lead to baryon
decay.

We can now choose the charges of the other Higgs fields in our
model, by considering the charge differences between
left-handed and right-handed fermions with the
inclusion of the WS Higgs. Since we have the constraint
of eq.~(\ref{diagU1}), the charges of these Higgs fields
must be chosen to span a 3 dimensional vector
space of charges represented, for example,
by $\frac{y_1}{2}$, $\frac{y_3}{2}$ and
$Q_f$ with $\frac{y_2}{2}$ being determined by
eq.~(\ref{diagU1}). This means that we will need at
least 3 Higgs fields to break the gauge
group down to the SMG. This gives us a lot of
freedom, so we will choose the charges on these Higgs
fields by considering phenomenological relations
between fermion masses.

Since we are assuming that the fundamental Yukawa
couplings are of order 1 but not exactly 1, we can
only produce order of magnitude results. So we wish to
choose, for example, 2 fermions with similar masses
but not order of magnitude equal masses.
We can then assume that the lighter fermion is suppressed
relative to the heavier fermion by 1 Higgs with a VEV
given approximately by the ratio of the 2 fermion masses.
For example we would say that the bottom quark and tau
lepton masses were of the same order of magnitude
(remembering that we take all relations at the Planck scale).
However we can take the following 2 ratios of effective
Yukawa couplings to be significantly different from 1:
\begin{equation}
\frac{g_c}{g_b} \equiv \frac{<W>}{M_F}
\approx \frac{1}{5}
\label{mc/mb}
\end{equation}
\begin{equation}
\frac{g_{\mu}}{g_b} \equiv \frac{<T>}{M_F}
\approx \frac{1}{13}
\label{mmu/mb}
\end{equation}
where we have defined 2 Higgs fields, $W$ and $T$,
to have appropriate VEVs, relative to the
fundamental mass scale $M_F$ of the theory, to cause
$m_c$ and $m_{\mu}$ to be suppressed relative to $m_b$.

First we define $\vec{b}$ to be the difference in charges
between $b_L$ and $b_R$ proto-fields
with the inclusion of the WS Higgs field. So we have:
\begin{equation}
\vec{b} = \vec{Q}_{b_L} - \vec{Q}_{b_R} - \vec{Q}_{WS}
\end{equation}
Similarly we define $\vec{c}$ and $\vec{\mu}$ to be:
\begin{eqnarray}
\vec{c} & = & \vec{Q}_{c_L} - \vec{Q}_{t_R} + \vec{Q}_{WS} \\
\vec{\mu} & = & \vec{Q}_{\mu_L} - \vec{Q}_{\mu_R} - \vec{Q}_{WS}
\end{eqnarray}
Note that $\vec{c}$ has been defined using the $t_R$
proto-field, since we have essentially
swapped the right-handed charm and top quarks.
Also the charges of the WS Higgs field
are added rather than subtracted for up-type quarks.
We observe that:
\begin{equation}
\vec{b}+\vec{c}+\vec{\mu}=\vec{0}
\label{b+c+mu=0}
\end{equation}
Now we can express these charges in terms of those of
the Higgs fields. We can define:
\begin{equation}
\vec{b} = \alpha\vec{Q}_W + \beta\vec{Q}_T + \vec{Q}_X
\end{equation}
where we have chosen the overall sign of the charges on the
Higgs fields $W$ and $T$ so that $\alpha$ and $\beta$
are not negative. $\vec{Q}_X$ is the total charges
of all other Higgs fields used to
suppress $m_b$ relative to $m_t$. We will assume
that $\vec{Q}_X$ cannot be expressed as a linear
combination of $\vec{Q}_W$
and $\vec{Q}_T$. Now eqs.~(\ref{mc/mb}) and (\ref{mmu/mb})
require that:
\begin{eqnarray}
\vec{c} & = & \pm (\alpha+1)\vec{Q}_W \pm
\beta\vec{Q}_T \pm \vec{Q}_X \label{cHiggs} \\
\vec{\mu} & = & \pm \alpha\vec{Q}_W \pm
(\beta+1)\vec{Q}_T \pm \vec{Q}_X \label{muHiggs}
\end{eqnarray}
The presence of the $\pm$ signs is due to the fact that
we can use the fields $W^{\dagger}$ and $T^{\dagger}$
as well as $W$ and $T$.

So we can rewrite eq.~(\ref{b+c+mu=0}) as:
\begin{equation}
\left( \begin{array}{c} 3\alpha+1 \\ \alpha+1 \\
\alpha-1 \\ -\alpha-1 \end{array}
\right) \vec{Q}_W +
\left( \begin{array}{c} 3\beta+1 \\ \beta-1 \\
\beta+1 \\ -\beta-1 \end{array}
\right) \vec{Q}_T +
\left( \begin{array}{c} 3 \\ 1 \\ 1 \\ -1 \end{array}
\right) \vec{Q}_X = \vec{0}
\label{QWTX}
\end{equation}
where the 4 coefficients for each term correspond to
the 4 combinations of signs in front of the terms
in eqs.~(\ref{cHiggs}) and (\ref{muHiggs}),
giving 64 cases altogether.

All possible choices of coefficient of $\vec{Q}_X$ are
non-zero and, by assumption, $\vec{Q}_X$ is linearly
independent of $\vec{Q}_W$ and $\vec{Q}_T$;
so eq.~(\ref{QWTX}) cannot hold.
We must
therefore conclude that there are no Higgs fields, other than
$W$ and $T$, used to
suppress $m_b$ relative to $m_t$. So we must set
$\vec{Q}_X=\vec{0}$. We can now use the fact that
$\alpha$ and $\beta$ are not negative, along
with the assumption that $\vec{Q}_T$ is not
directly proportional to $\vec{Q}_W$, to
conclude that eq.~(\ref{QWTX}) requires that:
\begin{equation}
\alpha=\beta=1
\end{equation}
and that the combination of signs is chosen so that:
\begin{eqnarray}
\vec{b} & = & \vec{Q}_W + \vec{Q}_T \\
\vec{c} & = & -2\vec{Q}_W + \vec{Q}_T \\
\vec{\mu} & = & \vec{Q}_W - 2\vec{Q}_T
\end{eqnarray}

We note that this immediately implies the reasonably
good Planck scale relation:
\begin{equation}
g_b = \; \frac{<W>}{M_F} \frac{<T>}{M_F} \;
\approx \frac{1}{65}
\end{equation}
This expression for the effective SM bottom quark Yukawa
coupling constant arises from the Feynman diagram in
figure 2. Here we have assumed the existence of a rich
spectrum of vector-like Dirac fermions, with unsuppressed
masses of the order of the fundamental mass scale
$M_F = M_{Planck}$, which provides the required
intermediate states. Also the fundamental Yukawa couplings
$\lambda_i$ are taken of order unity.

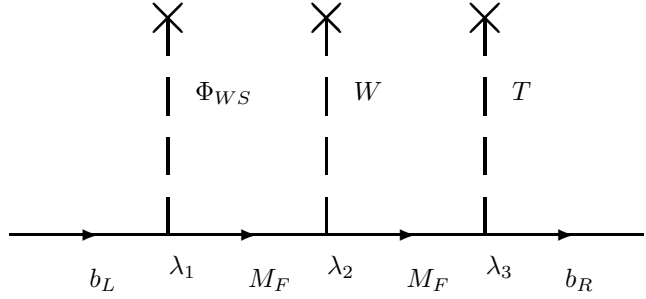
\begin{figure}
\begin{picture}(24000,11000)(-5600,0)
\THICKLINES

\drawline\fermion[\E\REG](0,1500)[6000]
\drawarrow[\E\ATBASE](\pmidx,\pmidy)
\global\advance \pmidy by -2000
\put(\pmidx,\pmidy){$b_L$}

\put(6000,0){$\lambda_1$}

\drawline\fermion[\E\REG](6000,1500)[6000]
\drawarrow[\E\ATBASE](\pmidx,\pmidy)
\global\advance \pmidy by -2000
\put(\pmidx,\pmidy){$M_F$}

\put(12000,0){$\lambda_2$}

\drawline\fermion[\E\REG](12000,1500)[6000]
\drawarrow[\E\ATBASE](\pmidx,\pmidy)
\global\advance \pmidy by -2000
\put(\pmidx,\pmidy){$M_F$}

\put(18000,0){$\lambda_3$}

\drawline\fermion[\E\REG](18000,1500)[6000]
\drawarrow[\E\ATBASE](\pmidx,\pmidy)
\global\advance \pmidy by -2000
\put(\pmidx,\pmidy){$b_R$}

\drawline\scalar[\N\REG](6000,1500)[4]
\global\advance \pmidx by 1000
\global\advance \pmidy by 1000
\put(\pmidx,\pmidy){$\Phi_{WS}$}
\global\advance \scalarbackx by -530
\global\advance \scalarbacky by -530
\drawline\fermion[\NE\REG](\scalarbackx,\scalarbacky)[1500]
\global\advance \scalarbacky by 1060
\drawline\fermion[\SE\REG](\scalarbackx,\scalarbacky)[1500]

\drawline\scalar[\N\REG](12000,1500)[4]
\global\advance \pmidx by 1000
\global\advance \pmidy by 1000
\put(\pmidx,\pmidy){$W$}
\global\advance \scalarbackx by -530
\global\advance \scalarbacky by -530
\drawline\fermion[\NE\REG](\scalarbackx,\scalarbacky)[1500]
\global\advance \scalarbacky by 1060
\drawline\fermion[\SE\REG](\scalarbackx,\scalarbacky)[1500]

\drawline\scalar[\N\REG](18000,1500)[4]
\global\advance \pmidx by 1000
\global\advance \pmidy by 1000
\put(\pmidx,\pmidy){$T$}
\global\advance \scalarbackx by -530
\global\advance \scalarbacky by -530
\drawline\fermion[\NE\REG](\scalarbackx,\scalarbacky)[1500]
\global\advance \scalarbacky by 1060
\drawline\fermion[\SE\REG](\scalarbackx,\scalarbacky)[1500]

\end{picture}
\caption{Feynman diagram for bottom quark mass in the AGUT
model. The crosses indicate the couplings of
the Higgs fields to the vacuum.}
\label{MbFull}
\end{figure}

It is now a simple matter to calculate the charges of the
Higgs fields $W$ and $T$.
We have:
\begin{equation}
\vec{Q}_W = \frac{1}{3}(2\vec{b}+\vec{\mu}) =
		\left( 0,-\frac{1}{2},\frac{1}{2},-\frac{4}{3} \right)
\end{equation}
{}From this we can then calculate:
\begin{equation}
\vec{Q}_T = \vec{b} - \vec{Q}_W =
\left( 0,-\frac{1}{6},\frac{1}{6},-\frac{2}{3} \right)
\end{equation}

We notice that the charges of $W$ and $T$ do not cover
the 2 dimensional space of charges $\frac{y_3}{2}$ and
$Q_f$, since only even $Q_f$ charges can be
constructed with integer numbers of these Higgs fields.
Therefore, since both $W$
and $T$ have $\frac{y_1}{2}=0$, we will
need at least 2 more Higgs fields to fully cover
the 3 dimensional charge space
required to break $G$ down to the SM
group. We will now choose 2
more Higgs fields which, together with $W$ and $T$,
will fully cover this space.

Another parameter in the SM, which is within one
order of magnitude from unity, is the mixing matrix
element between the 1st and 2nd generations,
which we associate with another Higgs field VEV:
\begin{equation}
V_{us} \equiv <\xi> \approx 0.2
\label{Vus}
\end{equation}
With the mass matrix texture in our model, $V_{us}$ is
approximately given by the ratio of the
mass matrix transition element from $d_L$ to
$s_R$ to the transition from $s_L$ to $s_R$.
This means that we must have:
\begin{equation}
\vec{Q}_{\xi} = \vec{Q}_{d_L} - \vec{Q}_{s_L}
	= \left( \frac{1}{6},0,0,0 \right) -
	\left( 0,\frac{1}{6},0,0 \right)
	= \left( \frac{1}{6},-\frac{1}{6},0,0 \right)
\end{equation}

{}From the well-known Fritzsch relation \cite{Fritschrule}
$V_{us} \simeq \sqrt{\frac{m_d}{m_s}}$,
it is suggested that the two off-diagonal mass
matrix elements connecting the
d-quark and the s-quark be equally big.
We achieve this approximately
in our model by introducing a special Higgs field
$S$, with quantum numbers equal to the
difference between the quantum number
differences for these 2 matrix elements in the
down quark matrix.
Then we postulate that this Higgs field has
a VEV of order unity in fundamental units,
so that it does not cause any suppression but
rather ensures that the two matrix elements
get equally suppressed. Henceforth we will
consider the VEVs of the new Higgs fields as
measured in units of $M_F$, and so we have:
\begin{equation}
<S> = 1
\end{equation}
and the charges of $S$ are given by:
\begin{eqnarray}
\vec{Q}_{S} & = & [\vec{Q}_{s_L} - \vec{Q}_{d_R}]
		- [\vec{Q}_{d_L} - \vec{Q}_{s_R}] \nonumber \\
 & = & \left[ \left( 0,\frac{1}{6},0,0 \right) -
		\left( -\frac{1}{3},0,0,0 \right) \right] -
	\left[ \left( \frac{1}{6},0,0,0 \right) -
		\left( 0,-\frac{1}{3},0,-1 \right) \right] \nonumber \\
 & = & \left( \frac{1}{6},-\frac{1}{6},0,-1 \right)
\end{eqnarray}
The existence of a non-suppressing
field $S$ means that we cannot
control phenomenologically when this $S$-field is used.
Thus the quantum numbers of the other
Higgs fields $W$, $T$, $\xi$ and $\phi_{WS}$
given above have only been determined modulo those
of the field $S$.

\section{Mass matrices, predictions}

We define the mass matrices
by considering the mass terms in the SM to be given by:
\begin{equation}
{\cal L}=Q_LM_uU_R+Q_LM_dD_R+L_LM_lE_R+{\rm h.c.}
\end{equation}
The mass matrices can be expressed in terms of the
effective SM Yukawa matrices and the WS Higgs VEV by:
\begin{equation}
M_f = Y_f \frac{<\phi_{WS}>}{\sqrt{2}}
\end{equation}
We can now calculate the suppression factors for
all elements in the Yukawa matrices, by expressing the
charge differences between the left-handed and
right-handed fermions in terms of the
charges of the Higgs fields. They are
given by products of the small numbers
denoting the VEVs, in fundamental units,
of the fields $W$, $T$, $\xi$ and
the of order unity VEV of $S$.
In the following matrices, we simply write $W$ instead of
$<W>$ etc. for the VEVs. With the quantum number
choice given above, the resulting matrix elements
are---but remember that ``random'' order
unity factors are supposed to multiply all the matrix
elements---for the uct-quarks:
\begin{equation}
Y_U \simeq \left ( \begin{array}{ccc}
	S^{\dagger}W^{\dagger}T^2(\xi^{\dagger})^2
	& W^{\dagger}T^2\xi & (W^{\dagger})^2T\xi \\
	S^{\dagger}W^{\dagger}T^2(\xi^{\dagger})^3
	& W^{\dagger}T^2 & (W^{\dagger})^2T \\
	S^{\dagger}(\xi^{\dagger})^3 & 1 & W^{\dagger}T^{\dagger}
			\end{array} \right ) \label{Y_U}
\end{equation}
the dsb-quarks:
\begin{equation}
Y_D \simeq \left ( \begin{array}{ccc}
	SW(T^{\dagger})^2\xi^2 & W(T^{\dagger})^2\xi & T^3\xi \\
	SW(T^{\dagger})^2\xi & W(T^{\dagger})^2 & T^3 \\
	SW^2(T^{\dagger})^4\xi & W^2(T^{\dagger})^4 & WT
			\end{array} \right ) \label{Y_D}
\end{equation}
and the charged leptons:
\begin{equation}
Y_E \simeq \left ( \hspace{-0.2 cm}\begin{array}{ccc}
	SW(T^{\dagger})^2\xi^2 & W(T^{\dagger})^2(\xi^{\dagger})^3
	& (S^{\dagger})^2WT^4\xi^{\dagger} \\
	SW(T^{\dagger})^2\xi^5 & W(T^{\dagger})^2 &
	(S^{\dagger})^2WT^4\xi^2 \\
	S^3W(T^{\dagger})^5\xi^3 & (W^{\dagger})^2T^4 & WT
			\end{array} \hspace{-0.2 cm}\right ) \label{Y_E}
\end{equation}

We can now set $S = 1$ and fit the nine quark and lepton masses
and three mixing angles, using 3 parameters: $W$, $T$
and $\xi$. That really means we have effectively omitted
the Higgs field $S$, and replaced the maximal AGUT gauge
group $SMG^3 \times U(1)_f$ by the reduced AGUT group
$SMG_{12} \times SMG_3 \times U(1)$, which survives the
spontaneous breakdown due to $S$.
In order to find the best possible fit, we
must use some function which measures how
good a fit is. Since we are expecting
an order of magnitude fit, this function
should depend only on the ratios of
the fitted masses to the experimentally
determined masses. The obvious choice
for such a function is:
\begin{equation}
\chi^2=\sum \left[\ln \left(
\frac{m}{m_{\mbox{\small{exp}}}} \right) \right]^2
\end{equation}
where $m$ are the fitted masses and mixing angles and
$m_{\mbox{\small{exp}}}$ are the
corresponding experimental values. The Yukawa
matrices are calculated at the fundamental scale,
which we take to be the
Planck scale. We use the first order renormalisation
group equations (RGEs) for
the SM to calculate the matrices at lower scales.

\begin{table}
\caption{Best fit to conventional experimental data.
All masses are running
masses at 1 GeV except the top quark mass
which is the pole mass.}
\begin{displaymath}
\begin{array}{ccc}
\hline
 & {\rm Fitted} & {\rm Experimental} \\ \hline
m_u & 3.6 {\rm \; MeV} & 4 {\rm \; MeV} \\
m_d & 7.0 {\rm \; MeV} & 9 {\rm \; MeV} \\
m_e & 0.87 {\rm \; MeV} & 0.5 {\rm \; MeV} \\
m_c & 1.02 {\rm \; GeV} & 1.4 {\rm \; GeV} \\
m_s & 400 {\rm \; MeV} & 200 {\rm \; MeV} \\
m_{\mu} & 88 {\rm \; MeV} & 105 {\rm \; MeV} \\
M_t & 192 {\rm \; GeV} & 180 {\rm \; GeV} \\
m_b & 8.3 {\rm \; GeV} & 6.3 {\rm \; GeV} \\
m_{\tau} & 1.27 {\rm \; GeV} & 1.78 {\rm \; GeV} \\
V_{us} & 0.18 & 0.22 \\
V_{cb} & 0.018 & 0.041 \\
V_{ub} & 0.0039 & 0.0035 \\ \hline
\end{array}
\end{displaymath}
\label{convbestfit}
\end{table}

We cannot simply use the 3 matrices given by
eqs.~(\ref{Y_U})--(\ref{Y_E}) to calculate
the masses and mixing angles, since
only the order of magnitude of the elements is defined.
Therefore we calculate
statistically, by giving each
element a random complex phase and then
finding the masses and mixing angles.
We repeat this several times and calculate
the geometrical mean
for each mass and mixing
angle. In fact we also vary the magnitude
of each element randomly, by
multiplying by a factor chosen to be
the exponential of a number picked from a
Gaussian distribution with mean value 0
and standard deviation 1.

We then vary the 3 free parameters to
find the best fit given by the $\chi^2$
function. We get the lowest value of $\chi^2$ for the VEVs:
\begin{eqnarray}
\langle W\rangle & = & 0.179   \label{Wvev} \\
\langle T\rangle & = & 0.071   \label{Tvev} \\
\langle \xi\rangle & = & 0.099 \label{xivev}
\end{eqnarray}
The fitted value of $\langle \xi\rangle$ is approximately
a factor of two smaller than the
estimate given in eq.~(\ref{Vus}).
This is mainly because there are
contributions to $V_{us}$ of the same
order of magnitude from both $Y_U$ and
$Y_D$. The result \cite{smg3m} of the fit is shown
in table~\ref{convbestfit}. This fit has a
value of:
\begin{equation}
\chi^2=1.87
\label{chisquared}
\end{equation}
This is equivalent to fitting 9 degrees of
freedom (9 masses + 3 mixing angles - 3
Higgs VEVs) to within a factor of
$\exp(\sqrt{1.87/9}) \simeq 1.58$
of the experimental value. It is
better than what one might have
expected from an order of magnitude
fit.

We can also fit to different experimental values
of the 3 light quark
masses by using recent results from lattice QCD, which
seem to be consistently lower than the conventional
phenomenological values.
The best fit in this case \cite{smg3m} is
shown in table~\ref{newbestfit}.
The values of the Higgs VEVs are:
\begin{eqnarray}
\langle W\rangle & = & 0.123	\\
\langle T\rangle & = & 0.079	\\
\langle \xi\rangle & = & 0.077
\end{eqnarray}
and this fit has a larger value of:
\begin{equation}
\chi^2 = 3.81
\end{equation}
But even this is good for an order of magnitude fit.

\begin{table}
\caption{Best fit using alternative light quark masses
extracted from lattice QCD. All masses are running
masses at 1 GeV except the top quark mass
which is the pole mass.}
\begin{displaymath}
\begin{array}{ccc}
\hline
 & {\rm Fitted} & {\rm Experimental} \\ \hline
m_u & 1.9 {\rm \; MeV} & 1.3 {\rm \; MeV} \\
m_d & 3.7 {\rm \; MeV} & 4.2 {\rm \; MeV} \\
m_e & 0.45 {\rm \; MeV} & 0.5 {\rm \; MeV} \\
m_c & 0.53 {\rm \; GeV} & 1.4 {\rm \; GeV} \\
m_s & 327 {\rm \; MeV} & 85 {\rm \; MeV} \\
m_{\mu} & 75 {\rm \; MeV} & 105 {\rm \; MeV} \\
M_t & 192 {\rm \; GeV} & 180 {\rm \; GeV} \\
m_b & 6.4 {\rm \; GeV} & 6.3 {\rm \; GeV} \\
m_{\tau} & 0.98 {\rm \; GeV} & 1.78 {\rm \; GeV} \\
V_{us} & 0.15 & 0.22 \\
V_{cb} & 0.033 & 0.041 \\
V_{ub} & 0.0054 & 0.0035 \\ \hline
\end{array}
\end{displaymath}
\label{newbestfit}
\end{table}

\section{ Baryogenesis}

A very important check of our model is whether or not
it can be consistent with baryogenesis.
In our model we have just the SM interactions up to about
one or two orders of magnitude under the Planck scale.
So we have no way, at the electroweak scale, to produce
the baryon number in the universe. There is insufficient
CP violation in the SM. Furthermore,
even if created, the baryon number
would immediately be washed out by sphaleron transitions
after the electroweak phase transition.
Our only chance to avoid
the baryon number being washed out at
the electroweak scale is to have a non-zero
$B-L$ (i.e. baryon number minus lepton number)
produced from the high, i.e. Planck, scale action
of the theory. That could then in turn give rise
to the baryon number at the electroweak scale.
Now, in our model, the $B-L$ quantum number
is broken by an anomaly
involving the $U(1)_f$ gauge group. This part of the
gauge group in turn  is broken by the Higgs field
$\xi$ which, in Planck units, is fitted to have
an expectation value around 1/10.
The anomaly keeps washing out any net $B-L$ that might appear,
due to CP-violating forces from the Planck scale physics,
until the temperature $T$ of the universe
has fallen to $\xi = 1/10$. The $U(1)_f$
gauge particle then disappears from the thermal soup and thus
the conservation of $B-L$ sets in.
The amount of $B-L$ produced at that time should then
be fixed and would
essentially make itself felt, at the electroweak scale,
by giving rise to an amount of
baryon number of the same order of magnitude.

The question now is whether we should
expect, in our model,
to have a
sufficient amount of time reversal
symmetry breaking, at the epoch
when $B-L$ settles down to be conserved, that the
amount of $B-L$ relative to the entropy (essentially
the amount of 3 degree Kelvin background radiation)
becomes large enough
to agree with the well-known phenomenological value of the
order of $10^{-9}$ or $10^{-10}$.
We shall use purely dimensional arguments, assuming
all couplings are generically of order unity, to
estimate the effects of Planck scale physics.
At the time of the order of the Planck scale,
when the temperature was also
of the order of the Planck temperature,
we expect even the CP or time reversal
violations were of order unity (in Planck units).
So at that time there existed
particles, say, with order of unity CP-violating decays.
However, they had also, in our purely dimensional
approximation, lifetimes
of the order of the Planck scale too.
Thus the $B-L$ biased decay products
would mainly be dumped at time 1 in Planck units, rather
than at the time of $B-L$ conservation setting in.
In a radiation dominated
universe, as we shall assume, the temperature
is proportional to 1/a where
a is the radius parameter---the size or scale
parameter of the universe.
Now the time goes as the square of this size parameter a.
Thus the time in Planck units is given
as the temperature to the negative second power:
\begin{equation}
t= \frac{0.3}{\sqrt{g_{\ast}} \times T^2}
\end{equation}
where \cite{utpal} $g_{\ast}$ is the number
of degrees of freedom---counted
as 1 for bosons but as 7/8 per
fermion degree of freedom---entering
into the radiation density. In our model $g_{\ast}$ gets a
contribution of $\frac{7}{8} \times 45 \times 2$
from the fermions and
$2 \times 37$ from the gauge bosons, and in addition
there is some contribution from the Higgs particles.
So we take $g_{\ast}$ to be of order 100, in our crude estimate of the
time t corresponding to the temperature
$T = \xi = \frac{1}{10}$
in Planck units, when $B-L$ conservation sets in:
\begin{equation}
t \simeq \frac{0.3}{\sqrt{100}} \times
\left(\frac{1}{10}\right)^{-2} = 3
\end{equation}
By that time we expect a fraction of the order of $\exp{(-3)}$
of the particles from the Planck era is still present and able
to dump its CP-violating
decay products. Of course here the uncertainty of, say,
an order of magnitude in $t$
would appear in the exponent, meaning a suppression factor
anywhere between $\exp{(0)}$ and $\exp{(-30)}$,
which could thus easily be in agreement with the value
wanted for baryogenesis of order
$5 \times 10^{-10}$. This result is encouraging,
but clearly a more careful analysis is required.

\section{Neutrino oscillations, a problem?}

We expect the neutrinos to get a mass in the AGUT model,
by the exchange of the WS Higgs field $\phi_{WS}$
twice---the see-saw mechanism \cite{fn2,seesaw}---or
the exchange of a weak isotriplet Higgs
field $\Delta$ \cite{gelmini}. In general
it will also be necessary
to exchange other AGUT Higgs scalars, in order to
balance the AGUT gauge quantum numbers beyond the SM.
This mechanism naturally generates an effective three
generation light neutrino mass matrix $M_{\nu}$:
\begin{equation}
{\cal L}_{m} = (M_{\nu})_{ij}\nu_{L_i}C\nu_{L_j} +\mbox{h.c.}
\label{Mnu}
\end{equation}
The neutrino mass matrix elements arise from Feynman
diagrams similar to those of figure 2, but they
involve two WS Higgs fields and the transitions are
between $\nu_{iL}$ and $\nu_{jR}^c$.
It follows that
\begin{equation}
M_{\nu} = H_{\nu}\frac{\langle \phi_{WS}\rangle^2}{2M_F}
\label{Mnuscale}
\end{equation}
where the dimensionless coupling matrix $H_{\nu}$ is
analogous to the quark-lepton Yukawa matrices $Y_U$,
$Y_D$ and $Y_E$, with elements expressed as products
of the AGUT Higgs field VEVs like in
eqs.~(\ref{Y_U})-(\ref{Y_E}).

Since the fundamental scale
of the AGUT model is
$M_F = M_{Planck}$,
the overall neutrino mass scale in the model, as
given by eq.~(\ref{Mnuscale}), is:
\begin{equation}
\frac{\langle \phi_{WS}\rangle^2}{2 M_{Planck}}
\sim 3 \times 10^{-6} \ \mbox{eV}
\end{equation}
This overall mass scale essentially provides an
upper limit to the neutrino masses. So
there is a basic problem for the AGUT model, since this
mass scale is too small to explain neutrino oscillation
phenomenology, except possibly for ``just-so'' vacuum
oscillations of solar neutrinos.

The neutrino mass matrix $M_{\nu}$ is, by its very definition
eq.~(\ref{Mnu}), symmetric. Also, in models like AGUT
with approximately conserved chiral $U(1)$ charges,
the matrix elements are generally of different orders
of magnitude, due to the presence of various suppression
factors. Thus the
generic structure for $M_{\nu}$ is a matrix in which the
various elements typically each have their own order
of magnitude, except in as far as they are forced to be equal
by the symmetry $M_{\nu} = M_{\nu}^T$.
The largest neutrino mass eigenvalue is then given by the
largest matrix element of $M_{\nu}$. If it
happens to be one of a pair of equal off-diagonal elements,
we get two very closely degenerate states as the heaviest
neutrinos with essentially maximal mixing; the
third neutrino will be much lighter and,
in first approximation, will not mix with the other two.
In fact this is what happens in the AGUT model, if we assume
that no new Higgs fields are introduced.

We can calculate the contributions to the neutrino mass
matrix from the Higgs fields already introduced
in our model and find:
\begin{equation}
H_{\nu} \sim \left ( \begin{array}{ccc}
	(S^{\dagger})^2(W^{\dagger})^2T^4(\xi^{\dagger})^4 &
		(S^{\dagger})^2(W^{\dagger})^2T^4\xi^{\dagger} &
		(W^{\dagger})^2T(\xi^{\dagger})^3 \\
	(S^{\dagger})^2(W^{\dagger})^2T^4\xi^{\dagger} &
		W(T^{\dagger})^5 & (W^{\dagger})^2T \\
	(W^{\dagger})^2T(\xi^{\dagger})^3 & (W^{\dagger})^2T &
		S^2(W^{\dagger})^2(T^{\dagger})^2(\xi^{\dagger})^2
			\end{array} \right )
\label{hnu}
\end{equation}
where, as usual, we assume that all fundamental
Yukawa couplings are of order 1.
Clearly all the elements of $H_{\nu}$ are suppressed.
The largest element is off-diagonal and of order
$\langle W\rangle^2\langle T\rangle$.  We
find the following eigenvalues for $H_{\nu}$:
\begin{eqnarray}
h_{\nu_\mu} \simeq h_{\nu_\tau} & \simeq &
\langle W\rangle^2\langle T\rangle  \simeq
2.3 \times 10^{-3} \\
h_{\nu_e} & \simeq & \langle W\rangle^2\langle
T\rangle^4\langle
\xi\rangle^4  \simeq 7.8 \times 10^{-11}
\end{eqnarray}
There is almost maximal mixing ($\sin^2 2\theta = 1$)
between $\nu_\mu$ and $\nu_\tau$ with
very small mixing ($\sin^2 2\theta \simeq \langle
\xi\rangle^6 \simeq 10^{-6}$) of $\nu_e$. This is not suitable
for observable solar
neutrino oscillations and, although we do have the correct
mixing structure for atmospheric neutrino oscillations, the masses
\begin{equation}
m_{\nu_{\mu}} \simeq m_{\nu_{\tau}} \simeq h_{\nu_{\tau}}
\frac{\langle \phi_{WS}\rangle^2}{2M_F} \simeq
7 \times 10^{-9} \ \mbox{eV}
\end{equation}
are far too small.

So we predict no observable neutrino oscillations, unless
we modify our AGUT model and introduce a new mass scale
into the theory. Either some intermediate mass see-saw
fermions or a weak isotriplet Higgs field
$\Delta$ is required. The latter could acquire a vacuum
expectation value of say
$\langle \Delta^0 \rangle \sim 1$ eV,
via its interaction with two WS Higgs
fields $\phi_{WS}$ and the other Higgs fields
$W$, $T$, $\xi$ and $S$; but only if,
for some as yet unknown reason,
it has a very small coefficient of $\Delta^2$ in the Higgs
potential compared to $M_{Planck}^2$.
Furthermore, it is difficult to explain both the solar
and atmospheric neutrino problems in the above scenario
where a single pair of equal off-diagonal elements dominates
$M_{\nu}$. Although it can provide the large
$\nu_{\mu}$ - $\nu_{\tau}$ mixing needed for the
atmospheric neutrino problem, their quasi-degeneracy in
mass implies that their mass differences with $\nu_e$
are too large to explain the solar neutrino
problem \cite{fgn}.
In fact it appears necessary for $M_{\nu}$ to
have at least two independent large elements being of
the same order of magnitude. Such an ``accidental''
order of magnitude degeneracy is perhaps not so
unlikely, in light of our discussion, in section 2, of the
number of order of magnitude degeneracies among the AGUT
charged fermion mass matrix elements.

\section{Conclusions}

We have tried to motivate the AGUT model from the
characteristic features of the quark-lepton masses and
mixing angles, which point rather strongly to proto-flavour
mass matrices having elements typically of different
orders of magnitude. We considered roughly how many
different orders of
magnitude would be represented by the proto-flavour mass matrices,
using experimental data and interpolating by theoretical
considerations. In this way, we estimated that the number of these
order of magnitude classes should be around 16. In turn
this estimate suggests around 7 to 8
cross product factors in the gauge group responsible
for the generation splittings etc.

The largest anomaly-free gauge group acting on just the
45 SM Weyl fermions, without any unification of the
SM irreducible representations, is the AGUT group
$SMG^3 \times U(1)_f$, which has 10 cross-product
factors giving 37 generators in all. This is broken
down to the reduced AGUT group $SMG^2 \times U(1)$
with 7 cross product factors
(which we actually use to fit the mass spectrum), by
the Higgs field $S$ with its VEV of order unity in
Planck units.

Now, inspired by the experimental data, we introduced three
more Higgs fields $W$, $T$ and $\xi$,
with appropriate quantum numbers made
to break the AGUT group down to the SM group.
We then presented a rather good fit, that in
principle should only work to order of magnitude
accuracy, to the charged quark-lepton masses and
mixing angles, using three parameters corresponding
to the VEVs of $W$, $T$ and $\xi$. The most characteristic
feature is that, apart from the $t$ and $c$ quarks,
the masses of the particles in the same generation are
predicted to be degenerate, but only order of
magnitude-wise, at the Planck scale. The worst feature of the
fit is the deviation by a factor of about 2 between the
fitted and experimental values for $m_s$ and $V_{cb}$.
However this is what can be expected in an order of
magnitude fit.

With this promising fit to the charged fermion masses,
we considered the predictions of the model for the baryon
number of the universe and neutrino masses.
Using a crude dimensional argument, together with the
expected number of degrees of freedom being of order 100,
we obtain results for baryogenesis consistent with the
observed baryon number, but with the uncertainty occurring
inside an exponent. If anything we tend to get too high a prediction
for the baryon number. However our predictions for the
neutrino masses are set by a Planck mass see-saw mechanism,
and are therefore too low to give observable neutrino
oscillations. This suggests we may have to modify
our model and introduce a new mass scale into the theory.

Apart from the neutrino puzzle, the AGUT model is
successful in explaining the quark-lepton masses and
mixing angles.

\end{document}